\documentclass[11pt,english]{extarticle}
\usepackage[T1]{fontenc}
\usepackage[latin9]{inputenc}
\usepackage[a4paper]{geometry}
\usepackage{bm}
\usepackage{amsmath}
\usepackage{amsthm}
\usepackage{amssymb}
\usepackage{setspace}
\usepackage[authoryear]{natbib}
\makeatletter
\theoremstyle{definition}
\newtheorem*{condition*}{\protect\conditionname}
\theoremstyle{plain}
\newtheorem{thm}{\protect\theoremname}
\newenvironment{lyxlist}[1]
	{\begin{list}{}
		{\settowidth{\labelwidth}{#1}
		 \setlength{\leftmargin}{\labelwidth}
		 \addtolength{\leftmargin}{\labelsep}
		 }}
	{\end{list}}
\theoremstyle{plain}
\newtheorem{lem}{\protect\lemmaname}

\@ifundefined{date}{}{\date{}}
\bibpunct{(}{)}{,}{a}{,}{,}

\usepackage{babel}

\usepackage[multiple]{footmisc}

\usepackage{chngcntr}
\usepackage{apptools}
\AtAppendix{\counterwithin{lemma}{section}}

\makeatletter
\def\thmhead@plain#1#2#3{%
  \thmname{#1}\thmnumber{\@ifnotempty{#1}{ }\@upn{#2}}%
  \thmnote{ {\the\thm@notefont#3}}}
\let\thmhead\thmhead@plain
\makeatother

\makeatother

\usepackage{babel}
\providecommand{\conditionname}{Condition}
\providecommand{\lemmaname}{Lemma}
\providecommand{\theoremname}{Theorem}

\begin{document}
\title{\vspace{20pt}
Subgeometric ergodicity and $\beta$-mixing\thanks{The authors thank the Academy of Finland for financial support. Contact
addresses: Mika Meitz, Department of Economics, University of Helsinki,
P. O. Box 17, FI\textendash 00014 University of Helsinki, Finland;
e-mail: mika.meitz@helsinki.fi. Pentti Saikkonen, Department of Mathematics
and Statistics, University of Helsinki, P. O. Box 68, FI\textendash 00014
University of Helsinki, Finland; e-mail: pentti.saikkonen@helsinki.fi.}\vspace{20pt}
}
\author{Mika Meitz\\\small{University of Helsinki} \and Pentti Saikkonen\\\small{University of Helsinki}\vspace{20pt}
}
\date{April 2019}
\maketitle
\begin{abstract}
\noindent It is well known that stationary geometrically ergodic
Markov chains are $\beta$-mixing (absolutely regular) with geometrically
decaying mixing coefficients. Furthermore, for initial distributions
other than the stationary one, geometric ergodicity implies $\beta$-mixing
under suitable moment assumptions. In this note we show that similar
results hold also for subgeometrically ergodic Markov chains. In particular,
for both stationary and other initial distributions, subgeometric
ergodicity implies $\beta$-mixing with subgeometrically decaying
mixing coefficients. Although this result is simple it should prove
very useful in obtaining rates of mixing in situations where geometric
ergodicity can not be established. To illustrate our results we derive
new subgeometric ergodicity and $\beta$-mixing results for the self-exciting
threshold autoregressive model.

\bigskip{}
\bigskip{}
\bigskip{}

\noindent \textbf{Classifications (MSC2010):} 60J05, 37A25.

\bigskip{}

\end{abstract}
\vfill{}
\pagebreak{}

\section{Introduction}

Let $X_{t}$ ($t=0,1,2,\ldots$) be a Markov chain on the state space
$\mathsf{X}$ with $n$-step transition probability measure $P^{n}$
and stationary distribution $\pi$. If the $n$-step probability measures
$P^{n}$ converge in total variation norm to the stationary probability
measure $\pi$ at rate $r^{n}$ (for some $r>1$), that is, 
\begin{equation}
\lim_{n\to\infty}r^{n}\lVert P^{n}(x\,;\,\cdot)-\pi(\cdot)\rVert=0,\quad\textrm{ \ensuremath{\pi\:}a.e.,}\label{eq:Geom-erg}
\end{equation}
the Markov chain is said to be geometrically ergodic. It is well known
that for stationary Markov chains, geometric ergodicity implies that
so-called $\beta$-mixing coefficients (or coefficients of absolute
regularity) $\beta(n)$, to be defined formally in Section 2, converge
to zero at the same rate, $\lim_{n\to\infty}r^{n}\beta(n)=0$ (see,
e.g., \citet[p. 89]{doukhan1994mixing}, \citet[Thm 3.7]{bradley2005basic},
or \citet[Thm 21.19]{bradley2007introduction}). For initial distributions
other than the stationary one, a similar mixing result has been obtained
by \citet[Propn 4]{liebscher2005towards}.

We are interested in counterparts of these mixing results when the
convergence in (\ref{eq:Geom-erg}) takes place at a rate $r(n)$
slower than geometric, that is, 
\begin{equation}
\lim_{n\to\infty}r(n)\lVert P^{n}(x\,;\,\cdot)-\pi(\cdot)\rVert=0,\quad\textrm{ \ensuremath{\pi\:}a.e..}\label{eq:SubGeom-erg}
\end{equation}
When (\ref{eq:SubGeom-erg}) holds with suitably defined rates $r(n)$
slower than geometric, the Markov chain is called subgeometrically
ergodic. The main result of this note establishes that for both stationary
and other initial distributions, subgeometric ergodicity implies $\beta$-mixing
with subgeometrically decaying mixing coefficients, that is, $\lim_{n\to\infty}\tilde{r}(n)\beta(n)=0$
for some rate function $\tilde{r}(n)$.

To illustrate some common rate functions, consider the expression
\[
r(n)=(1+\ln(n))^{\alpha}\,\cdot\,(1+n)^{\beta}\,\cdot\,e^{cn^{\gamma}}\,\cdot\,e^{dn},\qquad\alpha,\beta,c,d\geq0,\,\,\gamma\in(0,1),\,\,n\geq1.
\]
In the case $\alpha,\beta,c,d>0$ the four terms above satisfy $e^{dn}/e^{cn^{\gamma}}\to\infty$,
$e^{cn^{\gamma}}/(1+n)^{\beta}\to\infty$, and $(1+n)^{\beta}/(1+\ln(n))^{\alpha}\to\infty$
as $n\to\infty$, and this hierarchy can be used to define different
growth rates. Ordered from the fastest to the slowest growth rate,
a growth rate is called geometric (sometimes also exponential) if
the dominant term is $e^{dn}$ (with $d>0$; note that $e^{dn}=r^{n}$
with $r>1$ iff $d>0$), subexponential if the dominant term is $e^{cn^{\gamma}}$
($c>0$ and above $d=0$), polynomial if the dominant term is $(1+n)^{\beta}$
($\beta>0$, $c=d=0$), and logarithmic if the dominant term is $(1+\ln(n))^{\alpha}$
($\alpha>0$, $\beta=c=d=0$). 

To provide some brief background on subgeometric ergodicity, we note
that the first subgeometric ergodicity results for general state space
Markov chains were obtained by \citet{nummelin1983rate} and \citet{tweedie1983criteria};
the subgeometric rate functions $r(n)$ considered were introduced
by \citet{stone1967one}. \citet{tuominen1994subgeometric} gave a
set of conditions that imply the convergence in (\ref{eq:SubGeom-erg})
and, in particular, formulated a sequence of so-called drift conditions
to establish subgeometric ergodicity. Subsequent work by \citet{fort2000vsubgeometric},
\citet{jarner2002polynomial}, \citet{fort2003polynomial}, and \citet{douc2004practical}
lead to a formulation of a single drift condition to ensure subgeometric
ergodicity, paralleling the use of a Foster-Lyapunov drift condition
to establish geometric ergodicity (see, e.g., \citet[Ch 15]{meyn2009markov}).

The rest of the paper proceeds as follows. Section 2 contains necessary
mathematical preliminaries. Section 3 reviews the relation of geometric
ergodicity and $\beta$-mixing, while the corresponding results in
the subgeometric case are given in Section 4. The general results
obtained are exemplified in Section 5 where subgeometric ergodicity
and $\beta$-mixing results for the self-exciting threshold autoregressive
model are presented. Section 6 concludes, and all proofs are given
in an Appendix. 

\section{Preliminaries}

To formalize the discussion in the Introduction, consider $X_{t}$
($t=0,1,2,\ldots$), a time-homogeneous discrete-time Markov chain
on a general measurable state space $(\mathsf{X},\mathcal{B}(\mathsf{X}))$.
Comprehensive treatments of the relevant Markov chain theory can be
found in \citet{meyn2009markov} or \citet{douc2018markov}. Let $\mu$
be any initial measure on $\mathcal{B}(\mathsf{X})$, and suppose
that $X_{0}$ has distribution $\mu$. Denote the transition probabilities
with $P(x\,;\,A)$ ($x\in\mathsf{X}$, $A\in\mathcal{B}(\mathsf{X})$)
and let $(\Omega,\mathcal{F},\mathsf{P}_{\mu})$ denote the probability
space of the Markov process $\{X_{0},X_{1},\ldots\}$. As usual, $\mathsf{P}_{x}$
denotes the probability measure corresponding to a fixed initial value
$X_{0}=x$ and $P^{n}(x\,;\,A)=\mathsf{P}_{x}(X_{n}\in A)$ ($x\in\mathsf{X}$,
$A\in\mathcal{B}(\mathsf{X})$) signifies the $n$-step transition
probability measure. 

Next consider the rate of convergence of the $n$-step probability
measures $P^{n}$ to the stationary probability measure $\pi$. To
this end, for any two probability measures $\lambda_{1}$ and $\lambda_{2}$
on $(\mathsf{X},\mathcal{B}(\mathsf{X}))$, the total variation distance
is defined as $\lVert\lambda_{1}-\lambda_{2}\rVert=2\sup_{B\in\mathcal{B}(\mathsf{X})}\lvert\lambda_{1}(B)-\lambda_{2}(B)\rvert=\sup_{\lvert h\rvert\leq1}\lvert\lambda_{1}(h)-\lambda_{2}(h)\rvert$,
where the last supremum runs over all $\mathcal{B}(\mathsf{X})$-measurable
functions $h:\mathsf{X}\to\mathbb{R}$ bounded in absolute value by
1 and $\lambda_{i}(h)=\int_{\mathsf{X}}\lambda_{i}(dx)h(x)<\infty$.
The $n$-step probability measures $P^{n}$ converge in total variation
norm to the stationary probability measure $\pi$ at rate $r(n)$,
$n\geq0$, if 
\begin{equation}
\lim_{n\to\infty}r(n)\lVert P^{n}(x\,;\,\cdot)-\pi(\cdot)\rVert=0,\quad\textrm{ \ensuremath{\pi\:}a.e..}\label{f-ergodicity-1}
\end{equation}
If (\ref{f-ergodicity-1}) holds we say that the Markov chain $X_{t}$
is ergodic with rate $r(n)$; geometric ergodicity obtains when $r(n)=r^{n}$
for some $r>1$.

To define the $\beta$-mixing coefficients, let $\mathcal{F}_{k}^{l}$,
$0\leq k\leq l\leq\infty$, signify the $\sigma$-algebra generated
by $\{X_{k},\ldots,X_{l}\}$. For the stochastic process $\{X_{0},X_{1},\ldots\}$
the $\beta$-mixing coefficients $\beta(n)$, $n=1,2,\ldots$, are
defined as (\citet[Sec 1.1]{doukhan1994mixing}; \citet[Ch 3]{bradley2007introduction})
\begin{align*}
\beta(n) & =\frac{1}{2}\sup_{m\in\mathbb{N}}\sup\sum_{i=1}^{I}\sum_{j=1}^{J}\bigl|\mathsf{P}_{\mu}(A_{i}\cap B_{j})-\mathsf{P}_{\mu}(A_{i})\mathsf{P}_{\mu}(B_{j})\bigr|\\
 & =\sup_{m\in\mathbb{N}}\mathsf{E}_{\mu}\Bigl[\sup_{B\in\mathcal{F}_{n+m}^{\infty}}\bigl|\mathsf{P}_{\mu}(B\mid\mathcal{F}_{0}^{m})-\mathsf{P}_{\mu}(B)\bigr|\Bigr],
\end{align*}
where $\mathbb{N}=\{0,1,2,\ldots\}$ and in the first expression for
$\beta(n)$ the second supremum is taken over all pairs of (finite)
partitions $\{A_{1},A_{2},\ldots,A_{I}\}$ and $\{B_{1},B_{2},\ldots,B_{J}\}$
of $\Omega$ such that $A_{i}\in\mathcal{F}_{0}^{m}$ for each $i$
and $B_{j}\in\mathcal{F}_{n+m}^{\infty}$ for each $j$. For our purposes
it is convenient to use the following alternative expression obtained
by \citet[Propn 1; note that his definition of $\beta(n)$ includes an additional factor of 2]{davydov1973mixing}:
\begin{equation}
\beta(n)=\frac{1}{2}\sup_{m\in\mathbb{N}}\int_{\mathsf{X}}\mu P^{m}(dx)\left\Vert P^{n}(x\,;\,\cdot)-\mu P^{n+m}(\cdot)\right\Vert ,\quad n=1,2,\ldots,\label{Beta-mixing coeff}
\end{equation}
where $\mu P^{m}(\cdot)=\int_{\mathsf{X}}\mu(dx)P^{m}(x\,;\,\cdot)$
denotes the distribution of $X_{m}$ ($m=1,2,\ldots$; $\mu P^{0}=\mu$).
In case of a stationary Markov chain (i.e., one with initial distribution
$\pi$), the $\beta$-mixing coefficients can be expressed simply
as
\begin{equation}
\beta(n)=\frac{1}{2}\int_{\mathsf{X}}\pi(dx)\left\Vert P^{n}(x\,;\,\cdot)-\pi(\cdot)\right\Vert ,\quad n=1,2,\ldots.\label{beta-mixing coeff stat}
\end{equation}
Process $X_{t}$ is said to be $\beta$-mixing (or sometimes absolutely
regular) if $\lim_{n\rightarrow\infty}\beta(n)=0$. As with the convergence
in (\ref{f-ergodicity-1}), the rate of this convergence is of interest,
and in what follows we seek for results of the form $\lim_{n\rightarrow\infty}r(n)\beta(n)=0$
with some rate function $r(n)$.

\section{The geometric case}

We start by briefly discussing the relation of geometric ergodicity
and $\beta$-mixing; although these results are well known, comparing
them with the subgeometric case will be illuminating. In case of a
stationary Markov chain (i.e., one with initial distribution $\pi$),
this relation is particularly simple. As was first shown by \citet[Thm 2.1]{nummelin1982geometric},
a geometrically ergodic Markov chain satisfies, for some $r>1$, $\lim_{n\rightarrow\infty}r^{n}\int\pi(dx)\left\Vert P^{n}(x\,;\,\cdot)-\pi(\cdot)\right\Vert =0$;
given expression (\ref{beta-mixing coeff stat}), the $\beta$-mixing
property immediately follows and the mixing coefficients satisfy $\lim_{n\rightarrow\infty}r^{n}\beta(n)=0$.
Statements of this result can be found for instance in \citet[p. 89]{doukhan1994mixing},
\citet[Thm 3.7]{bradley2005basic}, and \citet[Thm 21.19]{bradley2007introduction}.
For initial distributions other than the stationary one, a corresponding
result seems to have first appeared in \citet[Propn 4]{liebscher2005towards}. 

To facilitate comparison with the subgeometric case, we present the
ergodicity and mixing results as consequences of a particular drift
criterion; as is discussed in \citet{meyn2009markov}, this is how
geometric ergodicity is often established. We use the following traditional
Foster-Lyapunov type geometric drift condition (cf.~\citet[Thm 15.0.1]{meyn2009markov}).\footnote{As a technical remark, note that in Condition Drift\textendash G we
assume the function $V$ to be everywhere finite (i.e., $V\,:\,\mathsf{X}\rightarrow[1,\infty)$)
and such that $\sup_{x\in C}V(x)<\infty$. In contrast, in \citet[Thm 15.0.1]{meyn2009markov}
it is only assumed that $V$ is extended-real-valued (i.e., $V\,:\,\mathsf{X}\rightarrow[1,\infty]$)
and finite at some one $x_{0}\in\mathsf{X}$. Our stronger requirements
hold in most practical applications and lead to more transparent exposition
and proofs.} Here $\boldsymbol{1}_{C}(\cdot)$ signifies the indicator function
of a set $C$. 
\begin{condition*}[\textbf{Drift\textendash G}]
Suppose there exist a petite set $C$, constants $b<\infty$, $\beta>0$,
and a measurable function $V\,:\,\mathsf{X}\rightarrow[1,\infty)$
such that $\sup_{x\in C}V(x)<\infty$, satisfying
\[
E\left[V(X_{1})\,\left|\,X{}_{0}=x\right.\right]\leq V(x)-\beta V(x)+b\boldsymbol{1}_{C}(x),\qquad x\in\mathsf{X}.
\]
\end{condition*}
For the definition of a `petite set' appearing in this condition,
and for the concepts of irreducibility and aperiodicity in the theorem
below, we refer the reader to \citet{meyn2009markov}. Theorem 1 summarizes
the relation between geometric ergodicity and $\beta$-mixing.
\begin{thm}
\noindent Suppose $X_{t}$ is a $\psi$-irreducible and aperiodic
Markov chain and that Condition Drift\textendash G holds. Then
\begin{lyxlist}{(x)}
\item [{\emph{(a)}}] $X_{t}$ is geometrically ergodic, i.e, for some $r_{1}>1$,
$\lim_{n\rightarrow\infty}r_{1}^{n}\left\Vert P^{n}(x\,;\,\cdot)-\pi(\cdot)\right\Vert =0$
for all $x\in\mathsf{X}$.
\end{lyxlist}
\noindent Suppose further that the initial state $X_{0}$ has distribution
$\mu$ such that $\int_{\mathsf{X}}\mu(dx)V(x)<\infty$. Then 
\begin{lyxlist}{(x)}
\item [{\emph{(b)}}] for some $r_{2}>1$, $\lim_{n\rightarrow\infty}r_{2}^{n}\int_{\mathsf{X}}\mu(dx)\left\Vert P^{n}(x\,;\,\cdot)-\pi(\cdot)\right\Vert =0$,
\end{lyxlist}
\noindent and 
\begin{lyxlist}{(x)}
\item [{\emph{(c)}}] $X_{t}$ is $\beta$-mixing and the mixing coefficients
satisfy, for some $r_{3}>1$, $\lim_{n\rightarrow\infty}r_{3}^{n}\beta(n)=0$.
\end{lyxlist}
\noindent Moreover:
\begin{lyxlist}{(x)}
\item [{\emph{(d)}}] In the stationary case ($\mu=\pi$) condition $\int_{\mathsf{X}}\pi(dx)V(x)<\infty$
is not needed, (b) and (c) hold with $r_{2}=r_{3}$, and (b) and (c)
are equivalent.
\end{lyxlist}
\end{thm}
Parts (a) and (b) are very well known (see for instance \citet[Thm 15.0.1]{meyn2009markov}
for part (a) and \citet[Thm 2.3]{nummelin1982geometric} for part
(b)) and so is also the mixing result in the stationary case (see
the references given earlier). Part (c) for general initial distributions
was obtained by \citet[Propn 4]{liebscher2005towards}, although our
formulation is somewhat different from his (our formulation and proof
avoid the use of so-called `$Q$-geometric ergodicity' employed by
Liebscher; for completeness, our proof of Theorem 1, which may be
of independent interest, is provided in a Supplementary Appendix).
Part (d) elaborates parts (b) and (c) as well as their relation in
the stationary case. 

\section{The subgeometric case}

We seek a counterpart of Theorem 1 in which the geometric rate $r^{n}$
is replaced by some slower rate function; such rate functions were
already exemplified in the Introduction. More formally, the subgeometric
rate functions we consider are defined as follows (cf., e.g., \citet{nummelin1983rate}
and \citet{douc2004practical}). Let $\Lambda_{0}$ be the set of
positive nondecreasing functions $r_{0}\,:\,\mathbb{N}\rightarrow[1,\infty)$
such that $\ln[r_{0}(n)]/n$ decreases to zero as $n\rightarrow\infty$.
The class of subgeometric rate functions, denoted by $\Lambda$, consists
of positive functions $r\,:\,\mathbb{N}\rightarrow(0,\infty)$ for
which there exists some $r_{0}\in\Lambda_{0}$ such that
\begin{equation}
0<\liminf_{n\rightarrow\infty}\frac{r(n)}{r_{0}(n)}\leq\limsup_{n\rightarrow\infty}\frac{r(n)}{r_{0}(n)}<\infty.\label{eq:SubG_RatesLims}
\end{equation}
Typical examples are obtained of rate functions $r$ for which these
inequalities hold with (for notational convenience, we set $\ln(0)=0$)
\[
r_{0}(n)=(1+\ln(n))^{\alpha}\,\cdot\,(1+n)^{\beta}\,\cdot\,e^{cn^{\gamma}},\qquad\alpha,\beta,c\geq0,\,\gamma\in(0,1).
\]
The rate function $r_{0}(n)$ is called subexponential when $c>0$,
polynomial when $c=0$ and $\beta>0$, and logarithmic when $\beta=c=0$
and $\alpha>0$.

In analogy with the geometric case, subgeometric ergodicity and mixing
results are most conveniently obtained by verifying an appropriate
drift condition. The following drift condition for subgeometric ergodicity
is adapted from \citet[Defn 16.1.7]{douc2018markov}.\footnote{A somewhat more general drift condition, for instance allowing for
$V$ to be extended-real-valued, is given in \citet{douc2004practical}.} 
\begin{condition*}[\textbf{Drift\textendash SubG}]
Suppose there exist a petite set $C$, a constant $b<\infty$, a
concave increasing continuously differentiable function $\phi\,:\,[1,\infty)\rightarrow(0,\infty)$
satisfying $\lim_{v\rightarrow\infty}\phi'(v)=0$, and a measurable
function $V\,:\,\mathsf{X}\rightarrow[1,\infty)$ such that $\sup_{x\in C}V(x)<\infty$
and
\[
E\left[V(X_{1})\,\left|\,X{}_{0}=x\right.\right]\leq V(x)-\phi(V(x))+b\boldsymbol{1}_{C}(x),\qquad x\in\mathsf{X}.
\]
\end{condition*}
Note that if $\phi(v)=\eta v$ ($\eta>0$), one obtains Condition
Drift\textendash G (but assumption $\lim_{v\rightarrow\infty}\phi'(v)=0$
rules this out; as we are interested in subgeometric rates of ergodicity,
assuming this means no loss of generality, see \citet[Remark 16.1.8]{douc2018markov}). 

Following \citet{douc2004practical} we next introduce a rate function,
denoted by $r_{\phi}$. First define the function $H_{\phi}(v)=\int_{1}^{v}\frac{dx}{\phi(x)}$,
where $\phi$ is as in Condition Drift\textendash SubG. The definition
implies that $H_{\phi}$ is a nondecreasing, concave, and differentiable
function on $[1,\infty)$, and it has an inverse $H_{\phi}^{-1}\,:\,[0,\infty)\rightarrow[1,\infty)$
which is increasing and differentiable (see \citet[Sec 2.1]{douc2004practical}).
Thus, we can define the rate function
\[
r_{\phi}(z)=(H_{\phi}^{-1})'(z)=\phi\circ H_{\phi}^{-1}(z).
\]
\citet[Lemma 2.3 and Proposition 2.5]{douc2004practical} show that
this rate function is subgeometric and that Condition Drift\textendash SubG
implies the convergence (\ref{f-ergodicity-1}) at rate $r_{\phi}(n)$. 

Theorem 2 summarizes the relation between subgeometric ergodicity
and $\beta$-mixing. Here $\lfloor k\rfloor$ denotes the integer
part of the real number $k$.
\begin{thm}
\noindent Suppose $X_{t}$ is a $\psi$-irreducible and aperiodic
Markov chain and that Condition Drift\textendash SubG holds. Then
\begin{lyxlist}{(x)}
\item [{\emph{(a)}}] $X_{t}$ is subgeometrically ergodic with rate $r_{\phi}(n)$,
i.e, $\lim_{n\rightarrow\infty}r_{\phi}(n)\left\Vert P^{n}(x\,;\,\cdot)-\pi(\cdot)\right\Vert =0$
for all $x\in\mathsf{X}$.
\end{lyxlist}
\noindent Suppose further that the initial state $X_{0}$ has distribution
$\mu$ such that $\int_{\mathsf{X}}\mu(dx)V(x)<\infty$. Then 
\begin{lyxlist}{(x)}
\item [{\emph{(b)}}] $\lim_{n\rightarrow\infty}r_{\phi}(n)\int\mu(dx)\left\Vert P^{n}(x\,;\,\cdot)-\pi(\cdot)\right\Vert =0$, 
\end{lyxlist}
\noindent and 
\begin{lyxlist}{(x)}
\item [{\emph{(c)}}] $X_{t}$ is $\beta$-mixing and the mixing coefficients
satisfy $\lim_{n\rightarrow\infty}\tilde{r}_{\phi}(n)\beta(n)=0$
for any rate function $\tilde{r}_{\phi}(n)$ such that $\limsup_{n\rightarrow\infty}\tilde{r}_{\phi}(n)/r_{\phi}(n_{1})<\infty$
where $n_{1}=\lfloor n/2\rfloor$.
\end{lyxlist}
Moreover:
\begin{lyxlist}{(x)}
\item [{\emph{(d)}}] In the stationary case ($\mu=\pi$) condition $\int_{\mathsf{X}}\pi(dx)V(x)<\infty$
is not needed, (b) and (c) hold with $r_{\phi}(n)=\tilde{r}_{\phi}(n)$,
and (b) and (c) (with $r_{\phi}(n)=\tilde{r}_{\phi}(n)$) are equivalent.
\item [{\emph{(e)}}] If $r_{\phi}(n)$ satisfies (\ref{eq:SubG_RatesLims})
with $r_{\phi,0}(n)=(1+\ln(n))^{\alpha}\cdot(1+n)^{\beta}\cdot e^{cn^{\gamma}}$
and $\tilde{r}_{\phi}(n)$ satisfies (\ref{eq:SubG_RatesLims}) with
$\tilde{r}_{\phi,0}(n)=(1+\ln(n))^{\alpha}\cdot(1+n)^{\beta}\cdot e^{\tilde{c}n^{\gamma}}$
for some $0<\tilde{c}<c2^{-\gamma}$, then $\limsup_{n\rightarrow\infty}\tilde{r}_{\phi}(n)/r_{\phi}(n_{1})<\infty$. 
\end{lyxlist}
\end{thm}
Of the results in Theorem 2, part (a) is given in Proposition 2.5
of \citet{douc2004practical}. Part (b) can be obtained by combining
Theorem 4.1 of \citet{tuominen1994subgeometric} and Proposition 2.5
of \citet{douc2004practical}, but in the proof we make use of the
work of \citet{nummelin1983rate}. Part (c) is new and illuminates
the relation between subgeometrically ergodic Markov chains and their
$\beta$-mixing properties, thereby providing a counterpart of a result
obtained by \citet[Propn 4]{liebscher2005towards} in the case of
geometric ergodicity. Part (d) is analogous to its counterpart in
Theorem 1 and provides further insight to parts (b) and (c) whereas
part (e) makes part (c) more concrete in the case of the most common
rate functions. For completeness, we give a detailed proof in the
Appendix.

As discussed in \citet[Sec 2.3]{douc2004practical} and \citet[Thm 1]{meitz2019subgear},
there is a connection between the function $\phi$ and the rate function
$r_{\phi}$, which can be used to find out the latter in particular
cases. For instance, polynomial rate functions are associated with
cases where the function $\phi$ is of the form $\phi(v)=cv^{\alpha}$
with $\alpha\in(0,1)$ and $c\in(0,1]$, and then the rate obtained
is $r_{\phi}(n)=n^{\alpha/(1-\alpha)}$ (an alternative form is $r_{\phi}(n)=n^{\kappa-1}$
with $\kappa=1+\alpha/(1-\alpha)$ already given by \citet{jarner2002polynomial}).
In the subexponential case the function $\phi$ is such that $v/\phi(v)$
goes to infinity slower than polynomially so that a possibility, given
in \citet[Thm 1]{meitz2019subgear}, is $\phi(v)=c(v+v_{0})/[\ln(v+v_{0})]^{\alpha}$
for some $\alpha,c,v_{0}>0$. This results in the rate $r_{\phi}(n)=(e^{d})^{n^{1/(1+\alpha)}}$
for some $d>0$ which is faster than polynomial. A logarithmic rate
is an example of a rate slower than polynomial. Then the function
$\phi$ is of the form $\phi(v)=c[1+\ln(v)]^{\alpha}$ for some $\alpha>0$
and $c\in(0,1]$, and the resulting rate is $r_{\phi}(n)=[\ln(v)]^{\alpha}$
(see \citet[Sec 2.3]{douc2004practical}).

Theorem 2 (or 1) also provides information about the moments of the
stationary distribution of $X_{t}$. Specifically, once part (a) of
Theorem 2 (or 1) has been established, one can deduce from Condition
Drift\textendash SubG (or Drift\textendash G) and Theorem 14.3.7 of
\citet{meyn2009markov} that $\int_{\mathsf{X}}\pi(dx)\phi(V(x))<\infty$
(or $\int_{\mathsf{X}}\pi(dx)V(x)<\infty$). This can be very useful
when one aims to apply limit theorems developed for $\beta$-mixing
processes where moment conditions are typically assumed.

We close this section by noting that Condition Drift\textendash SubG
can also be used to obtain more general ergodicity results than provided
in Theorem 2. Without going into details we only mention that Theorem
2.8 of \citet{douc2004practical} and Theorem 1 of \citet{meitz2019subgear}
show how a stronger form of ergodicity, called ($f,r$)-ergodicity,
can be established.

\section{Example}

To illustrate our results we consider the self-exciting threshold
autoregressive (SETAR) model studied by \citet{chan1985multiple}.
These authors analyzed the model 
\begin{equation}
X_{t}=\varphi(j)+\theta(j)X_{t-1}+W_{t}(j),\qquad X_{t-1}\in(r_{j-1},r_{j}],\label{SETAR}
\end{equation}
where $-\infty=r_{0}<\cdots<r_{M}=\infty$ and for each $j=1,\ldots,M$,
$\{W_{t}(j)\}$ is an independent and identically distributed mean
zero sequence independent of $\{W_{t}(i)\}$, $i\neq j$, and with
$W_{t}(j)$ having a density that is positive on the whole real line.
They considered the following conditions \begin{subequations}
\begin{align}
 & \theta(1)<1,\quad\theta(M)<1,\quad\theta(1)\theta(M)<1,\label{SETAR_params_a}\\
 & \theta(1)=1,\quad\theta(M)<1,\quad0<\varphi(1),\label{SETAR_params_b}\\
 & \theta(1)<1,\quad\theta(M)=1,\quad\varphi(M)<0,\label{SETAR_params_c}\\
 & \theta(1)=1,\quad\theta(M)=1,\quad\varphi(M)<0<\varphi(1),\label{SETAR_params_d}\\
 & \theta(1)<0,\quad\theta(1)\theta(M)=1,\quad\varphi(M)+\varphi(1)\theta(M)>0,\label{SETAR_params_e}
\end{align}
\end{subequations} and showed that the SETAR model is ergodic if
and only if one of the conditions (\ref{SETAR_params_a})\textendash (\ref{SETAR_params_e})
holds (\citealt[Thm 2.1]{chan1985multiple}). Moreover, if $E[\vert W_{t}(j)\vert]<\infty$
for each $j$, they showed that condition (\ref{SETAR_params_a})
ensures geometric ergodicity (\citealt[Thm 2.3]{chan1985multiple}).
To our knowledge, in the cases (\ref{SETAR_params_b})\textendash (\ref{SETAR_params_e})
no results regarding the rate of ergodicity have as yet appeared in
the literature and our Theorem 4(b) below indicates that geometric
ergodicity may not always hold without stronger assumptions.\footnote{\citet[Sec 11.4.3 and Sec B.2]{meyn2009markov} also discuss the (geometric)
ergodicity of the SETAR model (\ref{SETAR}), reproducing the ergodicity
result of \citet[Thm 2.1]{chan1985multiple} as their Proposition
11.4.5. On their p.~541, \citet{meyn2009markov} also state that
(our additions in brackets) ``\emph{in the interior of the parameter
space} {[}the union of (\ref{SETAR_params_a})\textendash (\ref{SETAR_params_e}){]}
\emph{we are able to identify geometric ergodicity in Proposition
11.4.5\ \ldots\ the stronger form} {[}geometric ergodicity{]} \emph{is
actually proved in that result}'' but no formal proof is given for
this statement. }

We consider rates of ergodicity and $\beta$-mixing in case (\ref{SETAR_params_d})
when the autoregressive coefficients $\theta(1)$ and $\theta(M)$
equal unity. For intuition, note that due to nonzero intercept terms
$\varphi(1)$ and $\varphi(M)$, both the first and the last regimes
exhibit nonstationary random walk type behavior with a drift. As the
intercept terms satisfy $\varphi(M)<0<\varphi(1)$, the drift is increasing
in the first regime and decreasing in the last regime. This feature
prevents the process $y_{t}$ from exploding to (plus or minus) infinity,
thereby providing intuition why ergodicity can hold true. It is noteworthy
that ergodicity is in no way dependent of the behavior of the process
in the middle regimes ($2,\ldots,M-1$) which can exhibit stationary,
random walk type (with or without drift), or even explosive behavior. 

In their results, \citet{chan1985multiple} allow for regime dependent
distributions for the error term $W_{t}(j)$. To obtain our results
for the case (\ref{SETAR_params_d}), we strengthen the assumptions
on the error term and, in particular, assume that the error distribution
is the same in each regime (this stronger assumption is needed to
apply the results mentioned in the proof of Theorem 3 below, and relaxing
it appears less than straightforward). To compensate, we obtain results
for a model more general than the SETAR model (\ref{SETAR}) with
(\ref{SETAR_params_d}). Specifically, we formulate our results in
terms of the general nonlinear autoregressive model
\begin{equation}
X_{t}=g(X_{t-1})+\varepsilon_{t},\qquad t=1,2,\ldots,\label{NLAR}
\end{equation}
where the function $g\,:\,\mathbb{R}\rightarrow\mathbb{R}$ and the
error term $\varepsilon_{t}$ satisfy the following conditions:
\begin{lyxlist}{0000}
\item [{(A1)}] $g$ is a measurable function with the property $\left|g(x)\right|\rightarrow\infty$
as $\left|x\right|\rightarrow\infty$ and such that there exist positive
constants $r$ and $M_{0}$ such that 
\[
\left|g(x)\right|\leq\left(1-r/\left|x\right|\right)\left|x\right|\quad\textrm{for }\left|x\right|\geq M_{0}\quad\textrm{and}\quad{\textstyle \sup_{\left|x\right|\leq M_{0}}}\left|g(x)\right|<\infty;
\]
\item [{(A2)}] $\{\varepsilon_{t},\,t=1,2,\ldots\}$ is a sequence of independent
and identically distributed mean zero random variables that is independent
of $X_{0}$ and the distribution of $\varepsilon_{1}$ has a (Lebesgue)
density that is bounded away from zero on compact subsets of $\mathbb{R}$.
\end{lyxlist}
Model (\ref{NLAR}) with conditions A1 and A2 is a special case of
models considered by \citet[Sec 2.2]{fort2003polynomial}, \citet[Sec 3.3]{douc2004practical},
and \citet[Secs 3--4]{meitz2019subgear}. These authors consider much
more general models but for clarity of presentation we have simplified
the model as much as possible while still being able to obtain results
for the SETAR model (\ref{SETAR}) with (\ref{SETAR_params_d}) (the
first two of the abovementioned papers consider a multivariate version
of (\ref{NLAR}), whereas the third one considers a higher-order generalization
of (\ref{NLAR}); the inequality constraint for the function $g$
in condition A1 is also more general in these papers where it is only
required that $\left|g(x)\right|\leq\left(1-r\left|x\right|^{-\rho}\right)\left|x\right|$
for some $0<\rho\leq2$).

The following Theorem establishes ergodicity and $\beta$-mixing results
for model (\ref{NLAR}) with varying rates of convergence. The proof
(in the Appendix) makes use of results in \citet{fort2003polynomial},
\citet{douc2004practical}, and \citet{meitz2019subgear} to obtain
rates of ergodicity, as well as Theorems 1 and 2 above to obtain rates
of $\beta$-mixing (only the subgeometric mixing results in parts
(b) and (c) are new).
\begin{thm}
Consider model (\ref{NLAR}) with conditions (A1) and (A2). 
\begin{lyxlist}{(x)}
\item [{\emph{(a)}}] If $E\bigl[e^{z_{0}\left|\varepsilon_{1}\right|}\bigr]<\infty$
for some $\ensuremath{z_{0}>0}$, then $X_{t}$ is geometrically ergodic
with convergence rate $r(n)=r_{1}^{n}$ for some $r_{1}>1$. Moreover,
if the initial state $X_{0}$ has a distribution such that $E[e^{z\left|X_{0}\right|}]<\infty$
for some $z>0$, then $X_{t}$ is also $\beta$-mixing and the mixing
coefficients satisfy, for some $r_{3}>1$, $\lim_{n\rightarrow\infty}r_{3}^{n}\beta(n)=0$.
\item [{\emph{(b)}}] If $E\bigl[e^{z_{0}\left|\varepsilon_{1}\right|^{\kappa_{0}}}\bigr]<\infty$
for some $\ensuremath{z_{0}>0}$ and $\ensuremath{\kappa_{0}\in(0,1)}$,
then $X_{t}$ is subexponentially ergodic with convergence rate $r(n)=(e^{c})^{n^{\kappa_{0}}}$
(for some $c>0$). Moreover, if the initial state $X_{0}$ has a distribution
such that $E[e^{z\left|X_{0}\right|^{\kappa_{0}}}]<\infty$ for some
$z>0$, then $X_{t}$ is also $\beta$-mixing and the mixing coefficients
satisfy, for some $\tilde{c}>0$, $\lim_{n\rightarrow\infty}(e^{\tilde{c}})^{n^{\kappa_{0}}}\beta(n)=0$.
\item [{\emph{(c)}}] If $E\left[\left|\varepsilon_{1}\right|^{s_{0}}\right]<\infty$
for either $s_{0}=2$ or $s_{0}\geq4$, then $X_{t}$ polynomially
ergodic with convergence rate $r(n)=n^{s_{0}-1}$. Moreover, if the
initial state $X_{0}$ has distribution such that $E\left[\left|X_{0}\right|^{s_{0}}\right]<\infty$,
then $X_{t}$ is also $\beta$-mixing and the mixing coefficients
satisfy $\lim_{n\rightarrow\infty}n^{s_{0}-1}\beta(n)=0$.
\end{lyxlist}
\end{thm}
Theorem 3 shows that there is a trade-off between rates of ergodicity
and $\beta$-mixing and finiteness of moments of the error term. The
fastest geometric rate is obtained when $E\bigl[e^{z_{0}\left|\varepsilon_{1}\right|}\bigr]<\infty$
($z_{0}>0$) so that $\varepsilon_{1}$ has finite moments of all
orders and the slowest polynomial rate is obtained when only $E\left[\varepsilon_{1}^{2}\right]<\infty$.
As discussed after Theorem 2, we also have $\int_{\mathsf{X}}\pi(dx)\phi(V(x))<\infty$
so that there is a similar trade-off between these convergence rates
and finiteness of moments of the stationary distribution (expressions
of $V$ and $\phi$ are available in the proof of Theorem 3).

Above it was mentioned that \citet{fort2003polynomial}, \citet{douc2004practical},
and \citet{meitz2019subgear} consider (subgeometric) ergodicity of
models more general than (\ref{NLAR}) with conditions (A1) and (A2).
Making use of our Theorems 1 and 2, subgeometric rates of $\beta$-mixing
can straightforwardly be obtained also for these more general models.
We omit the details for brevity. 

In a series of papers, Veretennikov and co-authors also considered
the model (\ref{NLAR}) with function $g$ satisfying $\left|g(x)\right|\leq\left(1-r\left|x\right|^{-\rho}\right)\left|x\right|$
for some $1\leq\rho\leq2$. Using methods very different from ours,
they obtained results on subgeometric ergodicity and subgeometric
rates for $\beta$-mixing coefficients. The cases $1<\rho<2$ and
$\rho=2$ are considered in \citet{veretennikov2000polynomial}, \citet{klokov2004sub,klokov2005subexponential},
and \citet{klokov2007lower} and are shown to lead to subgeometric
rates. For the case $\rho=1$ relevant for the SETAR example, these
papers refer to \citet{veretennikov1988bounds,veretennikov1991estimating}
and \citet{veretennikov1990rate}. A result corresponding to our Theorem
3(a) can be found in \citet[Thm 1]{veretennikov1990rate} but subgeometric
rates, such as those in our Theorem 3(b) and (c), do not seem to be
established in the case $\rho=1$.

We now specialize the results above to the SETAR model (\ref{SETAR})
with (\ref{SETAR_params_d}). It is easy to see that this model, with
the function $g$ in (\ref{NLAR}) defined as $g(x)=\sum_{j=1}^{M}[\varphi(j)+\theta(j)x]\bm{1}\{x\in(r_{j-1},r_{j}]\}$
(with $\bm{1}\{\cdot\}$ denoting the indicator function), satisfies
the condition in A1. Namely, for $x$ large enough and positive we
have $\lvert g(x)\rvert=g(x)=x+\varphi(M)=\lvert x\rvert-(-\varphi(M))$
whereas for $x$ small enough and negative we have $\lvert g(x)\rvert=-g(x)=-x-\varphi(1)=\lvert x\rvert-\varphi(1)$,
so that the inequality in A1 holds for $M_{0}>\max\{\left|r_{1}\right|,\left|r_{M-1}\right|\}$
and $r=\min\{\varphi(1),-\varphi(M)\}$ (and the supremum condition
is obviously satisfied). 

Part (a) of the next theorem simply restates the result of Theorem
3 for the SETAR model (\ref{SETAR}) with (\ref{SETAR_params_d}),
whereas part (b) establishes that geometric ergodicity cannot hold
under the weaker moment assumptions of Theorem 3(b) and (c). 
\begin{thm}
Consider the SETAR model (\ref{SETAR}) with the parameters satisfying
(\ref{SETAR_params_d}) and the error terms satisfying $W_{t}(j)=\varepsilon_{t}$
($j=1,\ldots,M$) with $\varepsilon_{t}$ as in (A2).
\begin{lyxlist}{(x)}
\item [{\emph{(a)}}] Sufficient conditions for geometric, subexponential,
and polynomial ergodicity and $\beta$-mixing of $X_{t}$ are as in
parts (a), (b), and (c) of Theorem 3, respectively.
\item [{\emph{(b)}}] If $E\bigl[e^{z_{0}\left|\varepsilon_{1}\right|}\bigr]=\infty$
for all $\ensuremath{z_{0}>0}$, then $X_{t}$ is not geometrically
ergodic.
\end{lyxlist}
\end{thm}
Theorem 4(b) shows that for the SETAR model (\ref{SETAR}) with (\ref{SETAR_params_d}),
the subgeometric rates of Theorem 3(b) and (c) cannot be improved
to a geometric rate unless stronger moment assumptions are made regarding
the error term. This result is obtained by making use of a necessary
condition for geometric ergodicity of certain specific type of Markov
chains in \citet{jarner2003necessary} (using their necessary condition
to obtain this result appears possible only in case (\ref{SETAR_params_d})
out of (\ref{SETAR_params_a})\textendash (\ref{SETAR_params_e})).

\section{Conclusion}

In this note we have shown that subgeometrically ergodic Markov chains
are $\beta$-mixing with subgeometrically decaying mixing coefficients.
Although this result is simple it should prove very useful in obtaining
rates of mixing in situations where geometric ergodicity can not be
established. An illustration using the popular self-exciting threshold
autoregressive model showed how our results can yield new subgeometric
rates of mixing. 

\pagebreak{}

\appendix

\section*{Appendix}

This Appendix contains the proofs of Theorems 2\textendash 4; proof
of Theorem 1 is provided in the Supplementary Appendix. Proofs of
Theorems 1 and 2 make use of the following handy inequality for the
$\beta$-mixing coefficients due to Liebscher (2005, Proposition 3).
(Note that our Lemma 1 below includes an additional factor of $\frac{1}{2}$
compared to Liebscher's Proposition 3; cf.~our expression for $\beta(n)$
in (\ref{Beta-mixing coeff}) and his eqn.~(27).) Again, $\lfloor k\rfloor$
denotes the integer part of the real number $k$. 
\begin{lem}
Suppose $X_{t}$ is a Markov chain with stationary distribution $\pi$
and that the initial state $X_{0}$ has distribution $\mu$. Then
\[
\beta(n)\leq\frac{1}{2}\int\pi(dx)\left\Vert P^{n_{1}}(x\,;\,\cdot)-\pi\right\Vert +\frac{3}{2}\int\mu(dx)\left\Vert P^{n_{1}}(x\,;\,\cdot)-\pi\right\Vert ,\quad n=1,2,\ldots,
\]
where $n_{1}=\lfloor n/2\rfloor$. 
\end{lem}
In the proof below, notation $\mathsf{E}_{\mu}\left[\cdot\right]$
is used for the conditional expectation of a $\mathcal{F}_{0}^{\infty}$-measurable
random variable conditioned on the initial state $X_{0}$ with distribution
$\mu$. When conditioning is on $X_{0}=x$ the notation $\mathsf{E}_{x}\left[\cdot\right]$
is used; these are connected via $\mathsf{E}_{\mu}\left[\cdot\right]=\int_{\mathsf{X}}\mu(dx)\mathsf{E}_{x}\left[\cdot\right]$.
We also define the concept of return time to a measurable set $A$
as $\tau_{A}=\inf\left\{ n\geq1:X_{n}\in A\right\} $. For brevity,
in the proof we refer to \citet{nummelin1983rate} and \citet*{douc2004practical}
as NT83 and DFMS04, respectively. 

\smallskip{}

\begin{proof}[\bfseries{\em Proof of Theorem 2}]
\textbf{} First note that, due to the assumed irreducibility and
aperiodicity, the petite set $C$ in Condition Drift\textendash SubG
is small (\citet[Thm 5.5.7]{meyn2009markov}). We first show that
\begin{equation}
\sup_{x\in C}\mathsf{E}_{x}\left[\sum_{k=0}^{\tau_{C}-1}\nolimits r_{\phi}(k)\right]<\infty;\label{eq:Th2_StopMoment1}
\end{equation}
by Theorem 2.1 of Tuominen and Tweedie (1994) this implies the subgeometric
ergodicity in (a) (for related results implying (a), see also NT83,
Theorem 2.7(i); \citet{tweedie1983criteria}, Theorem 1(iii); DFMS04
Proposition 2.5). It is shown in DFMS04 (Proposition 2.1 and Lemma
2.3) that Condition Drift\textendash SubG implies the existence of
a sequence of drift functions $V_{k}(x)$, $k=0,1,2,\ldots$, such
that, for $k\geq0$,
\[
E\left[V_{k+1}(X_{1})\,\left|\,X{}_{0}=x\right.\right]\leq V_{k}(x)-r_{\phi}(k)+\tilde{b}r_{\phi}(k)\boldsymbol{1}_{C}(x),
\]
where $\tilde{b}=br_{\phi}(1)(r_{\phi}(0))^{-2}$ (see their Proposition
2.1 and top of their page 1358) and $r_{\phi}\in\Lambda$ (see their
Lemma 2.3). Applying Proposition 11.3.2 of \citet{meyn2009markov}
with $Z_{k}=V_{k}(X_{k})$, $f_{k}(x)=r_{\phi}(k)$, $s_{k}(x)=\tilde{b}r_{\phi}(k)\boldsymbol{1}_{C}(x)$,
and stopping time $\tau_{C}$ we obtain (DFMS04, Proposition 2.2,
also states this conclusion; note also that by their eqn (2.2) we
have $V_{0}(x)\leq V(x)$)
\begin{equation}
\mathsf{E}_{x}\left[\sum_{k=0}^{\tau_{C}-1}\nolimits r_{\phi}(k)\right]\leq V(x)+\mathsf{E}_{x}\left[\sum_{k=0}^{\tau_{C}-1}\nolimits\tilde{b}r_{\phi}(k)\boldsymbol{1}_{C}(x)\right]=V(x)+\tilde{b}r_{\phi}(0)\boldsymbol{1}_{C}(x)=V(x)+b\frac{r_{\phi}(1)}{r_{\phi}(0)}\boldsymbol{1}_{C}(x).\label{eq:Th2_StopMoment2}
\end{equation}
By the condition $\sup_{x\in C}V(x)<\infty$ (in Condition Drift\textendash SubG),
we obtain (\ref{eq:Th2_StopMoment1}). Now, Theorem 2.1 of \citet{tuominen1994subgeometric}
ensures that $\lim_{n\to\infty}r_{\phi}(n)\left\Vert P^{n}(x\,;\,\cdot)-\pi(\cdot)\right\Vert =0$
so that the subgeometric ergodicity in (a) is established (note that
as $V_{0}(x)\leq V(x)$ holds with $V(x)$ assumed everywhere finite,
the set $S(f,r)$ in Theorem 2.1 of \citet{tuominen1994subgeometric}
coincides with $\mathsf{X}$ so that the aforementioned convergence
holds for all $x\in\mathsf{X}$). 

To prove (b), suppose the initial state $X_{0}$ has distribution
$\mu$ such that $\int_{\mathsf{X}}\mu(dx)V(x)<\infty$. We will use
Theorems 2.7(i,ii) and 2.2 of NT83, but first we obtain a property
of the rate function $r_{\phi}(z)$ (which is well-known for members
of $\Lambda_{0}$, but not for members of $\Lambda$). Recall that
$r_{\phi}(z)=(H_{\phi}^{-1})'(z)=\phi\circ H_{\phi}^{-1}(z)$ so that
$r_{\phi}'(z)/r_{\phi}(z)=\phi'\circ H_{\phi}^{-1}(z)$. As $\phi'$
is nonincreasing (see \citet[first paragraph of Sec 2.1]{douc2004practical})
and $H_{\phi}^{-1}$ is increasing, it follows that $r_{\phi}'(z)/r_{\phi}(z)=\phi'\circ H_{\phi}^{-1}(z)$
is nonincreasing. Therefore also the function $\ln(r_{\phi}(x))/x=\frac{1}{x}\int_{0}^{x}(r_{\phi}'(s)/r_{\phi}(s))ds$
($x>0$) is nonincreasing. Following the proof of Lemma 1 in \citet{stone1967one}
(which relies only on their property (iii) on their p.~326) yields
the desired property $r_{\phi}(m+n)\leq r_{\phi}(m)r_{\phi}(n)$ for
all $m,n>0$. 

Using this property we now obtain $r_{\phi}(\tau_{C})\leq r_{\phi}(1)r_{\phi}(\tau_{C}-1)\leq r_{\phi}(1)\sum_{k=0}^{\tau_{C}-1}r_{\phi}(k)$
and further $\mathsf{E}_{x}\bigl[\sum_{k=0}^{\tau_{C}}r_{\phi}(k)\bigr]\leq(r_{\phi}(1)+1)\mathsf{E}_{x}\bigl[\sum_{k=0}^{\tau_{C}-1}r_{\phi}(k)\bigr]$
and $\mathsf{E}_{x}\left[r_{\phi}(\tau_{C})\right]\leq r_{\phi}(1)\mathsf{E}_{x}\bigl[\sum_{k=0}^{\tau_{C}-1}r_{\phi}(k)\bigr]$
(cf.~Tuominen and Tweedie (1994, eqns (5) and (14)). The former result
together with (\ref{eq:Th2_StopMoment1}) implies that condition (2.12)
of Theorem 2.7(i) of NT83 is satisfied. The latter result together
with (\ref{eq:Th2_StopMoment2}) yields $\mathsf{E}_{x}\left[r_{\phi}(\tau_{C})\right]\leq r_{\phi}(1)[V(x)+b\frac{r_{\phi}(1)}{r_{\phi}(0)}\boldsymbol{1}_{C}(x)]$
and, as $\mathsf{E}_{\mu}\left[r_{\phi}(\tau_{C})\right]=\int_{\mathsf{X}}\mu(dx)\mathsf{E}_{x}\left[r_{\phi}(\tau_{C})\right]$,
the assumed bound $\int_{\mathsf{X}}\mu(dx)V(x)<\infty$ implies 
\begin{equation}
\mathsf{E}_{\mu}\left[r_{\phi}(\tau_{C})\right]<\infty,\label{eq:Th2_StopMoment3}
\end{equation}
so that the condition in Theorem 2.7(ii) of NT83 is satisfied. Therefore,
by Theorems 2.7(i,ii) and 2.2 of NT83, 
\[
\lim_{n\rightarrow\infty}r_{\phi}(n)\int\mu(dx)\left\Vert P^{n}(x\,;\,\cdot)-\pi(\cdot)\right\Vert =0.
\]

Next consider part (d). In the stationary case ($\mu=\pi$) the result
$\lim_{n\rightarrow\infty}r_{\phi}(n)\int\pi(dx)\left\Vert P^{n}(x\,;\,\cdot)-\pi(\cdot)\right\Vert =0$
follows from the last remark in Theorem 2.2 of NT83 (and condition
$\int_{\mathsf{X}}\pi(dx)V(x)<\infty$ is not needed). Thus (b) holds
in the stationary case. Regarding part (c) in the stationary case,
note from (\ref{beta-mixing coeff stat}) that now $\beta(n)=\int\pi(dx)\left\Vert P^{n}(x\,;\,\cdot)-\pi\right\Vert $,
$n=1,2,\ldots$, so that (b) and (c) are clearly equivalent (and hold
with the same rate $r_{\phi}(n)$). 

To prove (c), use Lemma 1 to obtain the inequality 
\[
\tilde{r}_{\phi}(n)\beta(n)\leq\frac{\tilde{r}_{\phi}(n)}{r_{\phi}(n_{1})}\left[\frac{1}{2}r_{\phi}(n_{1})\int\pi(dx)\left\Vert P^{n_{1}}(x\,;\,\cdot)-\pi\right\Vert +\frac{3}{2}r_{\phi}(n_{1})\int\mu(dx)\left\Vert P^{n_{1}}(x\,;\,\cdot)-\pi\right\Vert \right].
\]
The term in square brackets converges to zero as $n\to\infty$ by
parts (b) and (d) and, by assumption, $\limsup_{n\rightarrow\infty}\tilde{r}_{\phi}(n)/r_{\phi}(n_{1})<\infty$.
This establishes (c). 

To prove (e), it suffices to note that
\[
\frac{\tilde{r}_{\phi}(n)}{r_{\phi}(n_{1})}=\frac{\tilde{r}_{\phi}(n)}{\tilde{r}_{\phi,0}(n)}\frac{\tilde{r}_{\phi,0}(n)}{r_{\phi,0}(n_{1})}\frac{r_{\phi,0}(n_{1})}{r_{\phi}(n_{1})},
\]
where the first and the last ratio on the right hand side are bounded
from above uniformly in $n$ due to (\ref{eq:SubG_RatesLims}), and
that 
\[
r_{\phi,0}(n_{1})=\left(\frac{1+\ln(n_{1})}{1+\ln(n)}\right)^{\alpha}(1+\ln(n))^{\alpha}\,\cdot\,\left(\frac{1+n_{1}}{1+n}\right)^{\beta}(1+n)^{\beta}\,\cdot\,\frac{e^{cn_{1}^{\gamma}}}{e^{c(n/2)^{\gamma}}}e^{(c2^{-\gamma})n{}^{\gamma}},
\]
where the three ratios on the right hand side are clearly bounded
from below uniformly in $n$ by some constant larger than zero.
\end{proof}
\smallskip{}

\begin{proof}[\bfseries{\em Proof of Theorem 3}]
\textbf{} The ergodicity results of parts (a) and (b) could be obtained
using results in \citet[Sec 3.3]{douc2004practical} and those in
part (c) using results in \citet[Sec 2.2]{fort2003polynomial}; for
clarity of presentation, we will in all parts rely on the results
in \citet{meitz2019subgear}, henceworth MS19. Model (\ref{NLAR})
with conditions (A1) and (A2) is a special case of the model considered
in MS19 (with $p=\rho=1$ in that paper). Of the assumptions made
in MS19, Assumption 1 holds due to A1 and either Assumption 2(a) or
2(b) holds due to A2 and the moment conditions assumed in parts (a)\textendash (c)
of Theorem 3. Therefore we can make use Theorems 2 and 3 in MS19 to
obtain suitable ergodicity results. 

(a) In this case Assumption 2(a) of MS19 holds with $\kappa_{0}=1$
and we apply their Theorem 2(ii). From the proof of that theorem (Case
$p=1$) it can be seen that Condition Drift\textendash G holds with
$V(x)=e^{b_{1}\left|x\right|}$ for some $b_{1}\in(0,z_{0})$ which
can be chosen as small as desired. From Theorem 2(ii) of MS19 we obtain
that $X_{t}$ is geometrically ergodic with convergence rate $r(n)=(e^{c})^{n}$
(for some $c>0$), that is, $r(n)=r_{1}^{n}$ for some $r_{1}>1$.
To obtain results on $\beta$-mixing, we next apply Theorem 1 of the
present paper. If the initial state $X_{0}$ has distribution such
that $E[e^{z\left|X_{0}\right|}]<\infty$ for some $z>0$ (and noting
that above $b_{1}$ can be chosen small enough so that $b_{1}\leq z$
holds), then by Theorem 1 $X_{t}$ is $\beta$-mixing and the mixing
coefficients satisfy, for some $r_{3}>1$, $\lim_{n\rightarrow\infty}r_{3}^{n}\beta(n)=0$.

(b) In this case Assumption 2(a) of MS19 holds with $\kappa_{0}\in(0,1)$
and we apply their Theorem 2(i). From the proof of that theorem (Case
$p=1$) it can be seen that Condition Drift\textendash SubG holds
with $V(x)=e^{b_{1}\left|x\right|^{\kappa_{0}}}$ (for some $b_{1}\in(0,\beta_{0})$
which can be chosen as small as desired) and $\phi(v)=c_{0}(v+v_{0})(\ln(v+v_{0}))^{-\alpha}$
(for some $c_{0},v_{0}>0$ and $\alpha=1/\kappa_{0}-1$). From Theorem
2(i) of MS19 we obtain that $X_{t}$ is subexponentially ergodic with
convergence rate $r(n)=(e^{c})^{n^{\kappa_{0}}}$ (for some $c>0$).
To obtain results on $\beta$-mixing, we next apply Theorem 2 of the
present paper. If the initial state $X_{0}$ has distribution such
that $E[e^{z\left|X_{0}\right|^{\kappa_{0}}}]<\infty$ for some $z>0$
(and noting that above $b_{1}$ can be chosen small enough so that
$b_{1}\leq z$ holds), then by Theorem 2 $X_{t}$ is $\beta$-mixing
and the mixing coefficients satisfy, for any $\tilde{c}\in(0,z2^{-\kappa_{0}})$,
$\lim_{n\rightarrow\infty}\tilde{r}(n)\beta(n)=0$ with $\tilde{r}(n)=(e^{\tilde{c}})^{n^{\kappa_{0}}}$.

(c) In this case Assumption 2(b) of MS19 holds with either $s_{0}=2$
or $s_{0}\geq4$ and we apply their Theorem 3(ii) (in which exactly
the cases $s_{0}=2$ and $s_{0}\geq4$ are available). From the proof
of that theorem (the end of Step 4 and Case $p=1$) it can be seen
that Condition Drift\textendash SubG holds with $V(x)=1+\left|x\right|^{s_{0}}$
and $\phi(v)=cv^{\alpha}$ (for some $c>0$ and $\alpha=1-1/s_{0}$).
From Theorem 3(ii) of MS19 we obtain that $X_{t}$ is polynomially
ergodic with convergence rate $r(n)=n$ ($s_{0}=2$) or $r(n)=n^{s_{0}-1}$
($s_{0}\geq4$). To obtain results on $\beta$-mixing, we next apply
Theorem 2 of the present paper. If the initial state $X_{0}$ has
distribution such that $E\left[\left|X_{0}\right|^{s_{0}}\right]<\infty$,
then $X_{t}$ is $\beta$-mixing and the mixing coefficients satisfy
$\lim_{n\rightarrow\infty}n\beta(n)=0$ ($s_{0}=2$) or $\lim_{n\rightarrow\infty}n^{s_{0}-1}\beta(n)=0$
($s_{0}\geq4$).
\end{proof}
\smallskip{}

\begin{proof}[\bfseries{\em Proof of Theorem 4}]
\textbf{} Part (a) follows immediately from Theorem 3 and the discussion
preceding it noting that the SETAR model (\ref{SETAR}) with (\ref{SETAR_params_d})
satisfies the condition in A1. To prove (b), assume that $E\bigl[e^{z_{0}\left|\varepsilon_{1}\right|}\bigr]=\infty$
for all $\ensuremath{z_{0}>0}$ but that $X_{t}$ would be geometrically
ergodic. We will use results of \citet{jarner2003necessary} to show
that this leads to a contradiction. To this end, note that for the
SETAR model (\ref{SETAR}) with the parameters satisfying (\ref{SETAR_params_d})
the function $g$ in our equation (\ref{NLAR}) equals $g(x)=\sum_{j=1}^{M}[\varphi(j)+\theta(j)x]\bm{1}\{x\in(r_{j-1},r_{j}]\}$
which can be written as
\begin{align*}
g(x) & =[\varphi(1)+x]\bm{1}\{x\leq r_{1}\}+[\varphi(M)+x]\bm{1}\{r_{M-1}<x\}+{\textstyle \sum_{j=2}^{M-1}}[\varphi(j)+\theta(j)x]\bm{1}\{x\in(r_{j-1},r_{j}]\}\\
 & =x+\varphi(1)\bm{1}\{x\leq r_{1}\}+\varphi(M)\bm{1}\{r_{M-1}<x\}+{\textstyle \sum_{j=2}^{M-1}}[\varphi(j)+\theta(j)x-x]\bm{1}\{x\in(r_{j-1},r_{j}]\}
\end{align*}
or as $g(x)=x+\tilde{g}(x)$ where $\tilde{g}(x)$ is bounded. Also
recall that it is assumed that the error terms satisfy $W_{t}(j)=\varepsilon_{t}$
($j=1,\ldots,M$) with $\varepsilon_{t}$ as in (A2). These facts
show that the SETAR model (\ref{SETAR}) with (\ref{SETAR_params_d})
can be expressed in the form of equation (3) in \citet{jarner2003necessary}
so that $X_{t}$ is what \citet{jarner2003necessary} call a ``random-walk-type
Markov chain''. (Note also that this holds only in case (\ref{SETAR_params_d})
out of (\ref{SETAR_params_a})\textendash (\ref{SETAR_params_e}).)
Theorem 2.2 of \citet{jarner2003necessary} shows that a necessary
condition for the geometric ergodicity of a random-walk-type Markov
chain $X_{t}$ with stationary probability measure $\pi$ is that
there exists a $z>0$ such that $\int_{\mathsf{\mathbb{R}}}e^{z\left|x\right|}\pi(dx)<\infty$.
This can be shown to be in contradiction with our assumption that
$E\bigl[e^{z_{0}\left|\varepsilon_{1}\right|}\bigr]=\infty$ for all
$\ensuremath{z_{0}>0}$ as follows.

Suppose $z>0$ is such that $\int_{\mathsf{\mathbb{R}}}e^{z\left|x\right|}\pi(dx)<\infty$
and assume that $X_{0}$, and hence also $X_{1}$, has the stationary
distribution $\pi$. Thus $E[e^{z\left|X_{0}\right|}]<\infty$ and
$E[e^{z\left|X_{1}\right|}]<\infty$. As $0<e^{zx}\leq e^{z\left|x\right|}$
and $0<e^{-zx}\leq e^{z\left|x\right|}$, it follows that $E[e^{zX_{0}}]$,
$E[e^{-zX_{0}}]$, $E[e^{zX_{1}}]$, and $E[e^{-zX_{1}}]$ are all
positive and finite. As $X_{1}=X_{0}+\tilde{g}(X_{0})+\varepsilon_{1}$
with $X_{0}$ and $\varepsilon_{1}$ independent, $E[e^{zX_{1}}]=E[e^{zX_{0}}e^{z\tilde{g}(X_{0})}]E[e^{z\varepsilon_{1}}]$
(due to the nonnegativity of the exponential function, this holds
whether the expectations involved are finite or equal $+\infty$).
As $0<E[e^{zX_{0}}],E[e^{zX_{1}}]<\infty$ and $\tilde{g}(X_{0})$
is bounded this implies that $0<E[e^{z\varepsilon_{1}}]<\infty$.
An analogous argument yields that $0<E[e^{-z\varepsilon_{1}}]<\infty$.
Finally, nonnegativity of the random variables involved implies that
$E[e^{z\left|\varepsilon_{1}\right|}]=E[e^{z\varepsilon_{1}}\bm{1}\{\varepsilon_{1}\geq0\}+e^{-z\varepsilon_{1}}\bm{1}\{\varepsilon_{1}<0\}]\leq E[e^{z\varepsilon_{1}}]+E[e^{-z\varepsilon_{1}}]<\infty$,
yielding a contradiction.
\end{proof}
\smallskip{}
\bibliographystyle{chicago}
\bibliography{SubGE}

\pagebreak{}

\begingroup\singlespacing

\section*{Supplementary Appendix~~~~~{\small (not meant for publication)}}
\begin{proof}[\bfseries{\em Proof of Theorem 1}]
\textbf{}For brevity, we refer to \citet{nummelin1982geometric}
and \citet{meyn2009markov} as NT82 and MT09, respectively. We use
Theorem 2.5(ii) of NT82 to prove (a). To this end, first note that,
due to the assumed irreducibility and aperiodicity, the petite set
$C$ in Condition Drift\textendash G is small (MT09, Theorem 5.5.7).
We first show that, for some $r>1$, 
\[
{\textstyle \sup_{x\in C}}\mathsf{E}_{x}\left[r^{\tau_{C}}\right]<\infty;
\]
cf.~Theorem 2.5(ii) of NT82. We proceed as in the proof of Theorem
15.2.5 in MT09 and, for the $\beta$ in Condition Drift\textendash G,
choose an $r\in(1,(1-\beta)^{-1})$ and set $\varepsilon=r^{-1}-(1-\beta)$
so that $0<\varepsilon<\beta$ and $\varepsilon$ is the solution
to $r=(1-\beta+\varepsilon)^{-1}$. Now we may reorganize the drift
condition as 
\[
E\left[V(X_{1})\,\left|\,X{}_{0}=x\right.\right]\leq r^{-1}V(x)-\varepsilon V(x)+b\boldsymbol{1}_{C}(x),\qquad x\in\mathsf{X}.
\]
Define $Z_{k}=r^{k}V(X_{k})$, $k=0,1,2,\ldots$, so that $E[Z_{k+1}\mid\mathcal{F}_{0}^{k}]=r^{k+1}E[V(X_{k+1})\mid\mathcal{F}_{0}^{k}]$
and thus
\[
E[Z_{k+1}\mid\mathcal{F}_{0}^{k}]\leq r^{k+1}\{r^{-1}V(X_{k})-\varepsilon V(X_{k})+b\boldsymbol{1}_{C}(X_{k})\}=Z_{k}-\varepsilon r^{k+1}V(X_{k})+r^{k+1}b\boldsymbol{1}_{C}(X_{k}).
\]
Applying Proposition 11.3.2 of MT09 with $f_{k}(x)=\varepsilon r^{k+1}V(x)$,
$s_{k}(x)=br^{k+1}\boldsymbol{1}_{C}(x)$, and stopping time $\tau_{C}$
we obtain
\[
\mathsf{E}_{x}\left[\sum_{k=0}^{\tau_{C}-1}\nolimits\varepsilon r^{k+1}V(X_{k})\right]\leq V(x)+\mathsf{E}_{x}\left[\sum_{k=0}^{\tau_{C}-1}\nolimits br^{k+1}\boldsymbol{1}_{C}(X_{k})\right]=V(x)+br\boldsymbol{1}_{C}(x),
\]
because $\boldsymbol{1}_{C}(X_{1})=\cdots=\boldsymbol{1}_{C}(X_{\tau_{C}-1})=0$
by the definition of $\tau_{C}$. Multiplying by $\varepsilon^{-1}r^{-1}$
and noting that $V(\cdot)\geq1$, we obtain, for some finite constants
$c_{1},c_{2}$,
\[
\mathsf{E}_{x}\left[\sum_{k=0}^{\tau_{C}-1}\nolimits r^{k}\right]\leq\mathsf{E}_{x}\left[\sum_{k=0}^{\tau_{C}-1}\nolimits r^{k}V(X_{k})\right]\leq c_{1}V(x)+c_{2}.
\]
As $\sup_{x\in C}V(x)<\infty$, $\sup_{x\in C}\mathsf{E}_{x}\bigl[\sum_{k=0}^{\tau_{C}-1}r^{k}\bigr]<\infty$
is obtained. Using $\sum_{k=0}^{\tau_{C}-1}r^{k}=(r^{\tau_{C}}-1)/(r-1)$,
this is equivalent to $\sup_{x\in C}\mathsf{E}_{x}\left[r^{\tau_{C}}\right]<\infty$
as desired; note that we also have, for some finite constants $c_{3},c_{4}$,
$\mathsf{E}_{x}\left[r^{\tau_{C}}\right]\leq c_{3}V(x)+c_{4}$. Now
Theorem 2.5(ii) of NT82 implies that, for some $r_{1}>1$, $\lim_{n\rightarrow\infty}r_{1}^{n}\left\Vert P^{n}(x\,;\,\cdot)-\pi(\cdot)\right\Vert =0$,
so that the geometric ergodicity of part (a) is established.

To prove (b), suppose the initial state $X_{0}$ has distribution
$\mu$ such that $\int_{\mathsf{X}}\mu(dx)V(x)<\infty$. By Theorem
2.5(iii) of NT82 it suffices to prove that $\mathsf{E}_{\mu}\left[r^{\tau_{C}}\right]<\infty$.
As $\mathsf{E}_{\mu}\left[r^{\tau_{C}}\right]=\int_{\mathsf{X}}\mu(dx)\mathsf{E}_{x}\left[r^{\tau_{C}}\right]$,
the inequality $\mathsf{E}_{x}\left[r^{\tau_{C}}\right]\leq c_{3}V(x)+c_{4}$
obtained above implies $\mathsf{E}_{\mu}\left[r^{\tau_{C}}\right]<\infty$
and hence the validity of (b) for some $r_{2}>1$ (Theorem 2.5(iii)
of NT82). 

Next consider part (d). In the stationary case ($\mu=\pi$), the geometric
ergodicity established in (a) and Theorem 2.1 of NT82 imply that $\lim_{n\rightarrow\infty}\tilde{r}_{2}^{n}\int\pi(dx)\left\Vert P^{n}(x\,;\,\cdot)-\pi(\cdot)\right\Vert =0$
for some $\tilde{r}_{2}>1$ (and condition $\int_{\mathsf{X}}\pi(dx)V(x)<\infty$
is not needed). Thus (b) holds in the stationary case. Regarding part
(c) in the stationary case, note from (\ref{beta-mixing coeff stat})
that now $\beta(n)=\int\pi(dx)\left\Vert P^{n}(x\,;\,\cdot)-\pi(\cdot)\right\Vert $,
$n=1,2,\ldots$, so that (b) and (c) are clearly equivalent (and hold
with the same rate $\tilde{r}_{2}$). 

To prove (c) in the general case, recall that $n_{1}=\lfloor n/2\rfloor$
so that $n/2-1<n_{1}\leq n/2,$ and note that for any $\rho>1$ and
$n\geq2$, $1=\rho^{1-n/2}\rho^{n/2-1}<\rho^{1-n/2}\rho^{n_{1}}=\rho(\rho^{1/2})^{-n}\rho^{n_{1}}$.
Now choose $r_{3}$ such that $1<r_{3}<\min\{r_{2}^{1/2},\tilde{r}_{2}^{1/2}\}$
(where $r_{2}$ and $\tilde{r}_{2}$ are as above in the proofs of
parts (b) and (d)). Now use these remarks and the inequality in Lemma
1 to obtain 
\[
r_{3}^{n}\beta(n)\leq\frac{1}{2}\tilde{r}_{2}(r_{3}\tilde{r}_{2}^{-1/2})^{n}\tilde{r}_{2}^{n_{1}}\int\pi(dx)\left\Vert P^{n_{1}}(x\,;\,\cdot)-\pi\right\Vert +\frac{3}{2}r_{2}(r_{3}r_{2}^{-1/2})^{n}r_{2}^{n_{1}}\int\mu(dx)\left\Vert P^{n_{1}}(x\,;\,\cdot)-\pi\right\Vert .
\]
From the proofs of (b) and (d) we obtain the results $\lim_{n\rightarrow\infty}r_{2}^{n_{1}}\int\mu(dx)\left\Vert P^{n_{1}}(x\,;\,\cdot)-\pi\right\Vert =0$
and $\lim_{n\rightarrow\infty}\tilde{r}_{2}^{n_{1}}\int\pi(dx)\left\Vert P^{n_{1}}(x\,;\,\cdot)-\pi\right\Vert =0$,
so that $\lim_{n\rightarrow\infty}r_{3}^{n}\beta(n)=0$ and hence
(c) follows. 
\end{proof}
\endgroup
\end{document}